\documentclass[12pt, draftclsnofoot, onecolumn]{IEEEtran}
\usepackage[utf8]{inputenc}
\usepackage{graphicx}
\usepackage{amsmath}
\usepackage{amssymb}
\usepackage{amsfonts}
\usepackage{mathtools}
\usepackage{epsfig}
\usepackage{epstopdf}

\title{Adversarial Attack on DL-based Massive MIMO CSI Feedback}
\author{Qing~Liu,~Jiajia~Guo,~Chao-Kai~Wen,~Shi~Jin
	
	\thanks{Q. Liu is with the School of Cyber Science and Engineering, Southeast University, Nanjing
		210096, China (e-mail: qingliu@seu.edu.cn);
		
		 J. Guo and S. Jin are with the National
		Mobile Communications Research Laboratory, Southeast University, Nanjing
		210096, China (e-mail: jiajiaguo@seu.edu.cn; jinshi@seu.edu.cn);
	   
	    C.-K. Wen is with the Institute of Communications Engineering, National Sun Yat-sen University, Kaohsiung 804, Taiwan (e-mail: chaokai.wen@mail.nsysu.edu.tw).}
}	
\vspace{-3cm}
\date{}
\begin{document}

\maketitle
\begin{abstract}
    With the increasing application of deep learning (DL) algorithms in wireless communications, the physical layer faces new challenges caused by adversarial attack. Such attack has significantly affected the neural network in computer vision. We chose DL-based analog channel state information (CSI) to show the effect of adversarial attack on DL-based communication system. We present a practical method to craft white-box adversarial attack on DL-based CSI feedback process. Our simulation results showed the destructive effect adversarial attack caused on DL-based CSI feedback by analyzing the performance of normalized mean square error. We also launched a jamming attack for comparison and found that the jamming attack could be prevented with certain precautions. As DL algorithm becomes the trend in developing wireless communication, this work raises concerns regarding the security in the use of DL-based algorithms. 
\end{abstract}
\section{Introduction}
Deep learning (DL) is a promising technology in the Sixth Generation (6G) communication system \cite{1}. DL-based algorithm is used to deal with the huge data produced in the massive multiple-input multiple-output (MIMO) system. The DL-based algorithm can effectively optimize end-to-end performance despite the need for pre-defined mathematic models \cite{2,3}. As the DL-based model shows superiorities, the model is applied to physical layer broadly, e.g., channel estimation, modulation recognition, and channel state information (CSI) feedback \cite{4,5,6}.

Although DL is used increasingly, natural fragility makes it susceptible to adversarial attack in computer vision. Authors in \cite{7} presented a fast gradient method (FGM) to generate adversarial examples, which could lead to misclassification in neural network (NN)-based image classifier. DL becomes increasingly popular in wireless communication. Thus, attention is paid to the design of the network to improve transmission rate, whereas little attention is given to  DL-based physical layer security. Ref \cite{8} first indicated that DL-based modulation recognition network suffers from adversarial attack via FGM. The same research team futher improved FGM while launching an adversarial attack on an autoencoder-based end-to-end communication system in \cite{9}.

Although FGM is limited in launching adversarial attack on classification task, attack on the reconstruction task demands for a different design of adversarial perturbation. Inspired by \cite{10} that adversarial attack endangers DL-based classifier and maliciously leads to false feature extraction of image, we explore the threats as a possible adversarial attack against DL-based CSI feedback. The design is based on the similarity between feature extraction in computer vision and information compression in communication system.

In this paper, we study the security of the DL-based CSI feedback under adversarial attack. We launch a white-box adversarial attack on a DL-based massive MIMO CSI feedback, called CsiNet, which was proposed in \cite{6}. We then compare the output of CsiNet after the adversarial attack with the original input to evaluate the influence caused by the attack. We carry out a jamming attack as comparison. Our specific contributions are as follows. First, we discover that adversarial attack can cause a devastating impact on CSI feedback process compared to jamming attack. Second, we train the original model in AWGN channel and determine that training in this scenario can efficiently prevent CsiNet from Gaussian white noise. Nonetheless, adversarial attack could still disable CsiNet from proper functioning. Third, we compare the CsiNet trained with different compression rate and discover that the network with lower compression rate presents better robustness against adversarial attack. We conduct experiments in indoor and outdoor environments and find that the model suffers danger in different scenarios.

The remainder of this study is organized as follows. Section \uppercase\expandafter{\romannumeral2} introduces the system model and the CSI feedback that our experiments are based on. Section \uppercase\expandafter{\romannumeral3} gives a brief introduction of adversarial attack and proposes a method to attack CsiNet in detail. The simulation results and analysis are presented in Section \uppercase\expandafter{\romannumeral4}. The conclusion is given in Section \uppercase\expandafter{\romannumeral5}.

\section{System Model}

\subsection{Massive MIMO System}

We consider a system with \begin{math}N_t\end{math} antennas at the base station (BS) and a single antenna at user equipment (UE), which utilize orthogonal frequency division multiplexing (OFDM) with \begin{math}N_s\end{math} subcarriers. We denote the received signal of UE as follows:
\begin{equation}
	y_n=\widetilde{\mathbf{h}}_n^H \mathbf{v}_n x_n+ z_n
\end{equation}
where \begin{math}x_n\end{math} represents transmitted signal; \begin{math}z_n\end{math} is additional noise; \begin{math}\widetilde{\mathbf{h}}_n \in \mathbb{C}^{N_t \times 1}\end{math} and \begin{math}
\mathbf{v}_n\in \mathbb{C}^{N_t\times 1}\end{math} are channel frequency response and precoding vector, respectively. The downlink CSI can be described as \begin{math}\widetilde{\mathbf{H}}=[\widetilde{\mathbf{h}}_1,\widetilde{\mathbf{h}}_2,...,\widetilde{\mathbf{h}}_{N_s}]^H\in \mathbb{C}^{N_s\times N_t}\end{math} stacked in spatial frequency domain. To reach high quality downlink transmission, BS can design a channel precoding vector with knowledge of downlink CSI. Downlink CSI is first estimated at UE and then fed back to BS in frequency division dual (FDD) system. CSI feedback process can lead to a great overhead and occupy precious bandwidth. Therefore, 2D discrete Fourier transformation (DFT) is used to transform \begin{math}\widetilde{\mathbf{H}}\end{math} into angular-delay domain to reduce feedback overhead:
\begin{equation}
\mathbf{H}=\mathbf{F}_d  \widetilde{\mathbf{H}}  \mathbf{F}_a^H,
\end{equation}
where \begin{math}\mathbf{F}_d\end{math} and \begin{math}\mathbf{F}_a\end{math} are \begin{math}N_s\times N_s \end{math} and \begin{math}N_T\times N_T\end{math} DFT matrices.

As the time delay between multipath arrival lies within a limited period, the delay is presented such that the last several rows of \begin{math}\mathbf{H}\end{math} tend to be infinitely close to zero. Only the first \begin{math}N_c\end{math} rows of \begin{math}\mathbf{H}\end{math} exhibit non-zero values. Therefore, practical channel matrix \begin{math}\mathbf{H}\end{math} is truncated into \begin{math}N_c\times N_t\end{math}, by which the parameters waiting to be fed back are effectively reduced without interfering in transmission quality. 

\subsection{CSI Feedback}
Two main approaches of CSI feedback are available. One is digital CSI feedback, which compresses CSI, and quantization is applied to generate a bit stream for uplink transmission \cite{11}. The other is analog CSI feedback, which avoids quantization by transmitting downlink CSI through uplink channel using unquantized quadrature-amplitude modulation \cite{12}. Adversarial attack on digital CSI feedback is similar to attack on digital end-to-end communication system, which was proposed in \cite{9}. Therefore, we present a method to start adversarial attack on analog CSI feedback. Feedback CSI demands for a well-designed CSI sensing and recovery mechenism, which can be achieved by DL-based algorithm \cite{6}. 

Downlink CSI after 2D DFT can be visualized as an image where the gray-scale values represent the normalized absolute values of CSI. Since autoencoder has been proved to be an efficient model in dealing with image reconstruction, similarities between autoencoder and communication system are utilized by treating encoder as transmitter and decoder as receiver.

We denote an encoder by \begin{math} f_{\rm en}:\mathbf{H}\rightarrow \mathbb{C}^M \end{math}, where \begin{math}\mathbf{H}\end{math} refers to CSI matrix as input, and \begin{math}M\end{math} is the dimension of the encoder’s output. The encoder extracts latent representations from the original input. Then, representations would be reconstructed into the output with a decoder denoted by \begin{math} f_{\rm de}:\mathbf{s}\rightarrow \hat{\mathbf{H}} \end{math}, where \begin{math}\mathbf{s}=f_{\rm en}(\mathbf{H})\end{math} is an \begin{math}M\end{math} dimensional vector, which is the encoder output. Concatenating an encoder with a decoder forms an autoencoder, which would be trained to optimize end-to-end performance with the following loss function:
\begin{equation}
\min||f_{\rm de}(f_{\rm en}(\mathbf{H}))-\mathbf{H}||_2^2,
\end{equation}
where \begin{math}\mathbf{H}\end{math} refers to input.

\begin{figure*}[t]
	\centering
	\includegraphics[scale=0.6]{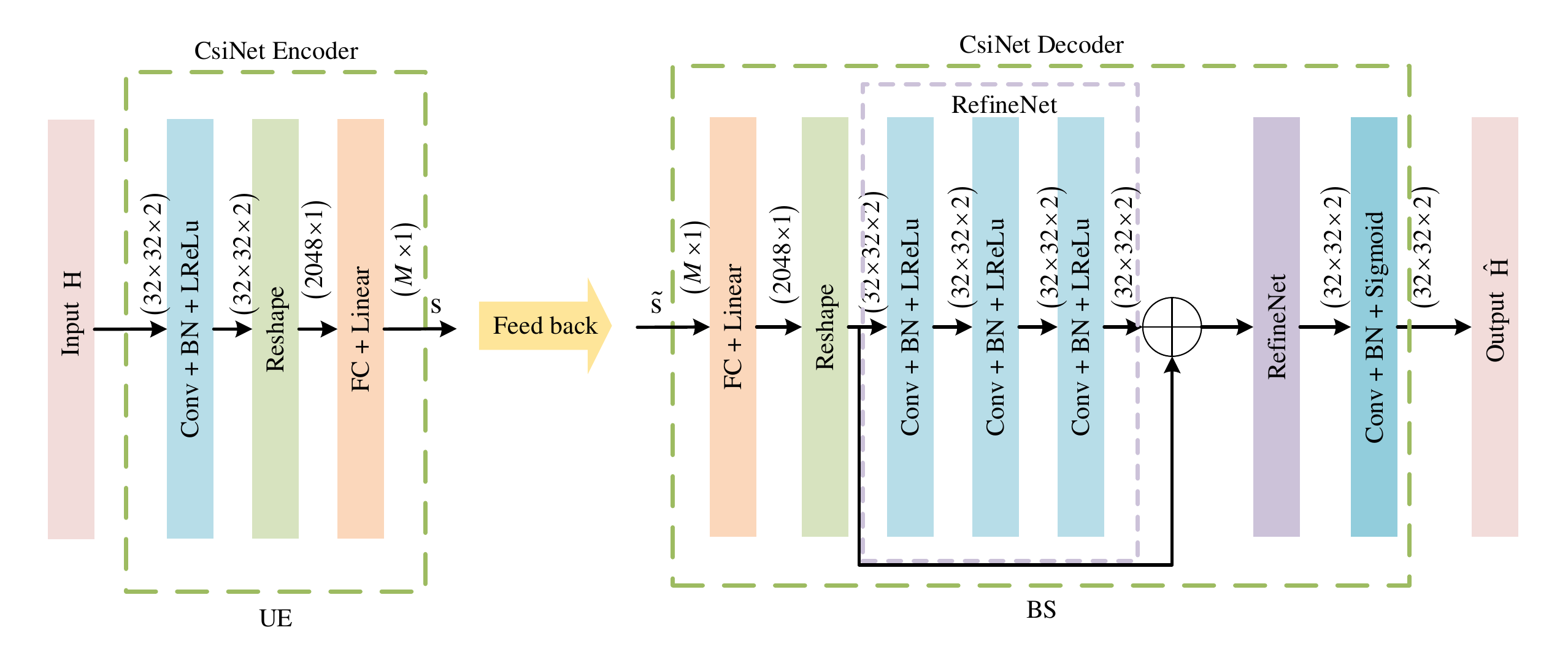}
	\caption{Architecture of CsiNet: An encoder constructed with convolutional, reshape, and fully connected layers; Decoder with fully connected and reshape layers and two RefineNet units connected in series. RefineNet unit is blocked specifically.}
	\label{fig:1}
\end{figure*}

Ref \cite{6} proposed an autoencoder-based NN, called CsiNet, to feedback downlink CSI. The structure of CsiNet is given in Fig. \ref{fig:1}, where the encoder placed at UE compresses CSI with one convolutional layer to generate two feature maps and a fully connected layer to generate two feature maps into an \begin{math}M\end{math} dimensional vector. Compression rate is defined as follows:
\begin{equation}
	\gamma=M/N,~{\rm with}~N=2 N_c\times N_t
\end{equation}
The decoder at BS is used to reconstruct original CSI from codewords. Codewords are fed into a fully connected layer to reverse the transformation made by the last layer of encoder. From the output of the fully connected layer, CSI is then refined by two RefineNet units connected in series and constructed by one last convolutional layer. The encoder and decoder would be trained jointly as a complete autoencoder to accomplish end-to-end optimization before placing seperately at UE and BS.

\section{Adversarial Attack on CsiNet}
In this section, we present the details of launching a white-box adversarial attack on CsiNet after a brief explanation of adversarial attack.

\subsection{Adversarial Attack in Theory}

First discovered in computer vision is that an intentionally imperceptible perturbation added on original input can generate an adversarial example, which could lead a misclassification with high confidence. The key to launching an adversarial attack is to find such adversarial perturbation, which can be fulfilled in theory by solving the following problem:
\begin{equation}
    \begin{split}
        & \displaystyle\max_{\mathbf{z}} ||f_{\rm de}(f_{\rm en}(\mathbf{H}+\mathbf{z}))-\mathbf{H}||_2^2 \\
        & s.t.~~||\mathbf{z}||_1\leq\delta,
    \end{split}
\end{equation}
where \begin{math}\mathbf{H}\end{math} refers to original input; \begin{math}\mathbf{z}\end{math} is an adversarial perturbation directly added on the input; and \begin{math}\delta\end{math} is set to constrain the adversarial perturbation. In computer vision, \begin{math}L_1\end{math} norm is used to limit perturbation to keep the variation away from human perception. The constraint could be changed into other forms, such as \begin{math}L_2\end{math} norm to fit other kinds of requirements. In wireless communication, \begin{math}L_2\end{math} norm is used to calculate signal power. Hence, we adopt \begin{math}L_2\end{math} norm to limit the power of perturbation. Moreover, (2) can be altered by adding a regularization part to keep the adversarial perturbation within limitations as follows:
\begin{equation}
\begin{split}
    & \displaystyle\max_{\mathbf{z}} ||f_{\rm de}(f_{\rm en}(\mathbf{H}+\mathbf{z}))-\mathbf{H}||_2^2+\epsilon||\mathbf{z}|| \\
    & s.t.~~||\mathbf{z}||_1\leq\delta,
\end{split}
\end{equation}
where \begin{math}\epsilon\end{math} is a scaling factor.  

Several ways can be applied to solve (5) or (6). According to \cite{7}, L-GBFS-b, which is a limited-memory algorithm for solving large nonlinear optimization problems subject to simple bounds on the variables \cite{13}, was presented to be a fine optimization in solving problems like (5). Ref \cite{7} proposed FGM to generate an adversarial perturbation, which leads a misclassification successfully. However, no specific classification is observed in attacking the reconstruction task, which makes FGM not feasible in generating an adversarial perturbation against autoencoders. The attack against autoencoders aims at the whole reconstruction. To find an adversarial perturbation against autoencoders, \cite{14} introduced that the parameters of addictive bias can make a well-functional perturbation by self-update of NN. We use their methods of generating perturbation in our experiments, which are introduced in the next subsection. 

\subsection{Adversarial Attack on CsiNet}
We propose a white-box adversarial attack on DL-based CSI feedback. The complete process of CSI feedback is accomplished by assuming that the perfect codeword is received by the BS. Due to the broadcast nature of wireless communication, we model an attacker, which could simultaneously send adversarial perturbation to be added to the transmitted data as follows:
\begin{equation}
\widetilde{\mathbf{s}}=\mathbf{s}+\mathbf{p},
\end{equation}
where \begin{math} \mathbf{s}=f_{\rm en}(\mathbf{H}) \end{math} denotes intact codewords; \begin{math} \mathbf{H} \end{math} is original CSI; and \begin{math}\mathbf{p}\end{math} denotes adversarial perturbation. Our goal is to generate a constant perturbation, which is added on transmitted codewords, and disable the decoder at BS. Subsequently, the BS would fail to reconstruct the perfect downlink CSI, which could further harm the communication system.

We model an attacker by a bias layer, which does the following calculation:
\begin{equation}
\mathbf{y}=g(\mathbf{h}+\mathbf{p}),
\end{equation} 
where we set activation \begin{math}g(\cdot)\end{math} to be linear, such that only the bias represented by \begin{math}\mathbf{p}\end{math} could be updated during back propagation.
 
We adopt a two-step training strategy by adding the bias layer between encoder and decoder of CsiNet. First, we initialize the parameters of bias layer to be zero, which are futher set to be non-trainable to train a functioning autoencoder designed to fulfill the task of CSI feedback using loss function as (3). After the model is trained, we set the model fixed and start to train the bias layer by using a loss function as follows:
\begin{equation}
\max||f_{\rm de}(\mathbf{s}+\mathbf{p})-\mathbf{H}||_2^2.
\end{equation}

Different from computer vision, the constraint we choose here cannot be imperceptible from human observation. To keep the power of perturbation within limitation, we previously set an interference-to-signal ratio (ISR), which is used to generate a value with power of codewords using the following equation:
\begin{equation}
||\mathbf{p}||_2^2/||\mathbf{s}||_2^2=\rm ISR.
\end{equation}
Hence, the parameters of the bias layer make an \begin{math}M\end{math} dimensional vector that can be used as the addictive adversarial perturbation to attack CsiNet. Epochs, number of training samples, and learning rates of each step of the training are given in Table \ref{tab:1}.
\begin{table}[t]
	\centering
	\vspace{-0.3cm}
	\caption{training parameters.}
	\label{tab:1}
	\begin{tabular}{c|c|c|c}
		\hline
		Step of training & epochs & training samples & learning rate \\
		\hline\hline
		first step & 200 & 100,000 & 0.001 \\
		\hline
		second step & 10 & 30,000 & 0.001 \\
		\hline
	\end{tabular}
	
\end{table}

After the two-step training is finished, we feed test data into the encoder to collect codewords, which are further added to a perturbation trained earlier. Tampered codewords are sent to the decoder to finish the reconstruction. We collect and compare the outputs of the decoder with the original CSI by calculating normalized mean square error (NMSE) between two elements as follows:
\begin{equation}
\rm NMSE=E\{||\mathbf{H}-\hat{\mathbf{H}}||_2^2/||\mathbf{H}||_2^2\},
\end{equation} 
where \begin{math}\mathbf{H}\end{math} denotes original CSI, and \begin{math}\hat{\mathbf{H}}\end{math} denotes the reconstructed CSI at BS. To evaluate the effect of adversarial attack, we launch a jamming attack on CsiNet to obtain outputs for the NMSE calculation. For jamming attack, we generate Gaussian white noise with the same power of adversarial perturbation to be added to the transmitted signal. We compare the NMSE performance of CsiNet under two kinds of attack for the same ISR and set the original NMSE performance of CsiNet without attack to be the baseline. 

CsiNet is first trained in an ideal scenario without consideration of noise. We then alter the training scenario by adding the AWGN channel between UE and BS. We study if the robustness of CsiNet against adversarial perturbation can be enhanced by certain precaution by adding random Gaussian noise with different power to codewords. In our experiments, we utilize signal-to-noise ratio (SNR) to set the power of Gaussian noise in the channel during training, where the signal refers to codewords \begin{math}\mathbf{s}\end{math}. We launch a jamming attack on newly trained CsiNet for comparison. Furthermore, we extend our experiments on CsiNet by adopting different compression rates. Considering that the practical scenario is complicated, we use indoor and outdoor CSI.

\section{Numerical Results}
We use COST 2100 channel model \cite{15} to generate two types of CSI dataset as indoor and outdoor scenarios for simulation. We set the carrier frequency indoor at 5.3 GHz and outdoor at 300 MHz. We place an ULA with 32 antennas at BS and a single antenna at UE in the OFDM system with 1024 subcarriers. Due to sparsity of massive MIMO-OFDM system, the practical complex channel matrix is truncated into \begin{math}32\times 32\end{math} after being transformed into an angular-delay domain. The training and validation dataset of the first-step training contain 100,000 and 30,000 samples, respectively. The training dataset of the second-step training contains 30,000 samples. An additional 20,000 samples are generated as testing dataset.
\begin{figure}[t]
	\centering
	\includegraphics[scale=0.63]{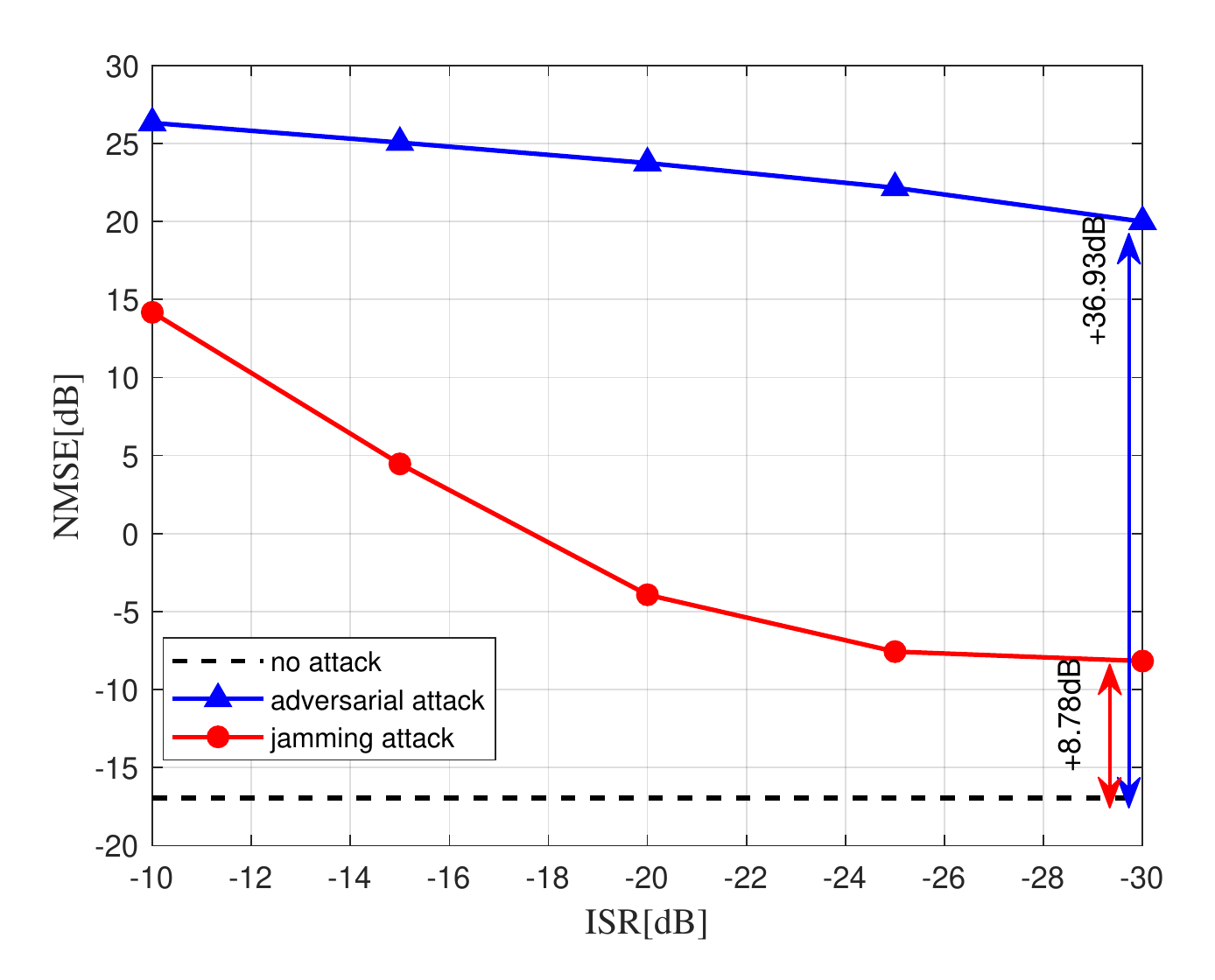}
	\caption{NMSE of CsiNet versus SNR with adversarial and jamming attacks for the indoor scenario with $\gamma$ set to be 1/4.}
	\label{fig:2}
\end{figure}

We first train CsiNet in an indoor ideal scenario without consideration of natural noise and with compression rate as \begin{math}1/4\end{math}. Adversarial and jamming attacks are launched in succession and NMSE performances of CsiNet under each kind of attacks are given in Fig. \ref{fig:2}. The horizontal dashed line represents the NMSE performance of CsiNet in the no attack scenario. From Fig. \ref{fig:2}, CsiNet presents a significantly higher NMSE while under adversarial attack compared to jamming attack. Hence, adversarial attack presents a more destructive influence on CsiNet for the same value of ISR. Moreover, the jamming attack is less of a threat when the power of noise drops. Adversarial attack holds a steadily destructive influence even when the power of perturbation is small.

\begin{figure}[t]
	\centering
	\includegraphics[scale=0.63]{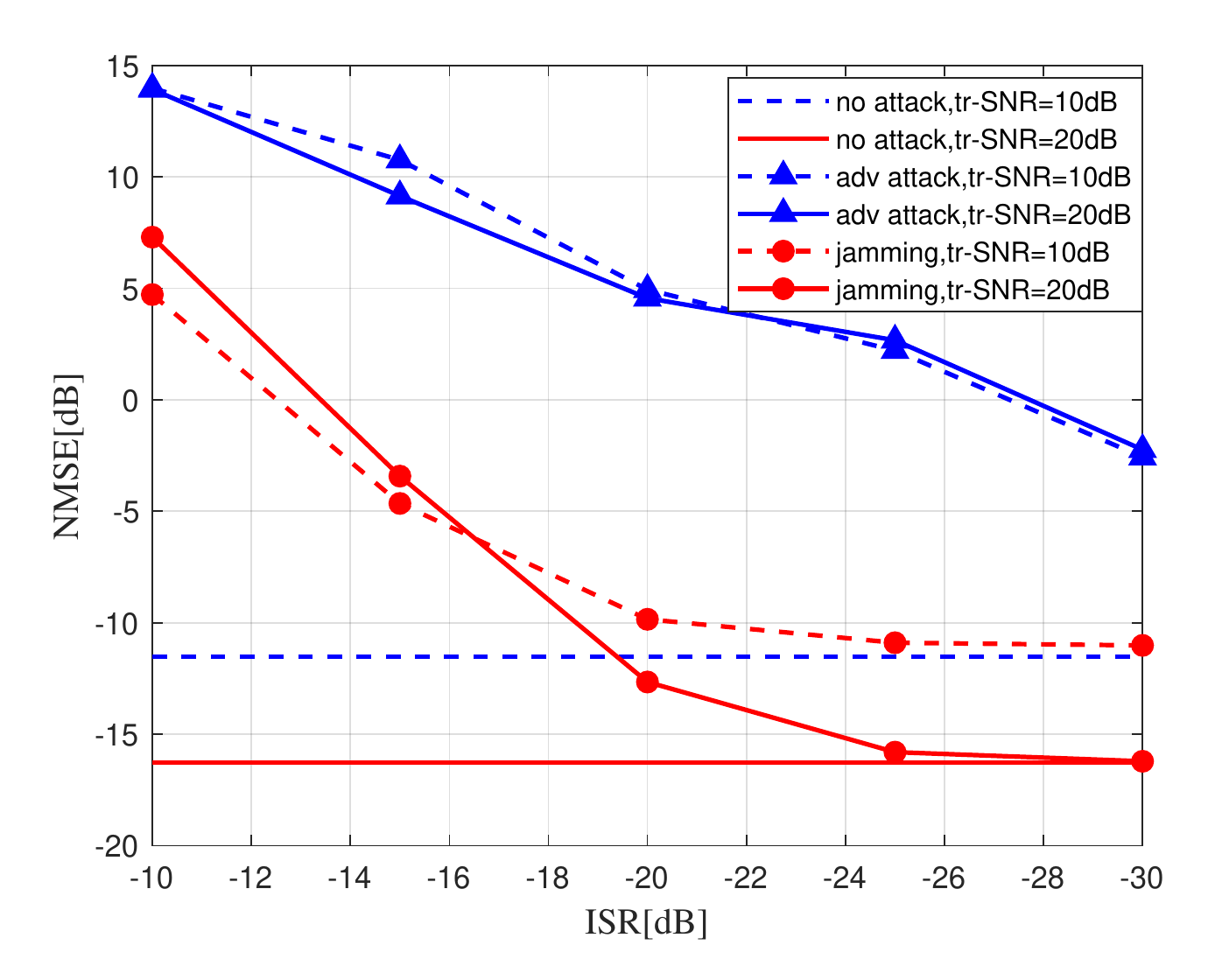}
	\caption{NMSE of CsiNet trained in AWGN channel versus SNR with adversarial attack for the indoor scenario with $\gamma$ set to be 1/4.}
	\label{fig:3}
\end{figure}

The results of Fig. \ref{fig:2} assume that BS could receive intact codewords. However, in wireless communication, channels are fragile because of natural noise. Therefore, blocks of physical layer should be designed with consideration of the complex channel state to enhance network robustness. Hence, we retrain the CsiNet in AWGN channel with different SNR set at 10 and 20 dB. Adversarial and jamming attacks are launched on the two new models whose results are given in Fig. \ref{fig:3}. To compare the NMSE performance of the new models under attack with previously trained models, we choose compression rate as \begin{math}1/4\end{math} of the indoor CSI dataset.

Fig. \ref{fig:3} shows that NMSE of the new model under jamming attack is infinitely closer to the baseline as the value of ISR drops. The NMSE performance of models trained in AWGN channel indicate that certain precaution can effectively enhance robustness of CsiNet against addictive Gaussian noise. However, Fig. \ref{fig:3} shows that NMSE performance of new models under adversarial attack drops slightly compared to Fig. \ref{fig:2}. However, the adversarial attack presents a high influence on CsiNet. When ISR is set to be \begin{math}-30\end{math} dB, NMSEs of CsiNet under adversarial attack are 8.94 dB and 14.03 dB higher than the baseline in 10 and 20 dB scenarios, respectively. In comparison with CsiNet trained without Gaussian noise, the new model shows slight resistance against adversarial attack and low-power adversarial perturbation still prevents CsiNet from proper functioning. Hence, the adversarial attack could effectively disable the CSI feedback despite precautionary measures. 

\begin{figure}[t]
	\centering
	\includegraphics[scale=0.63]{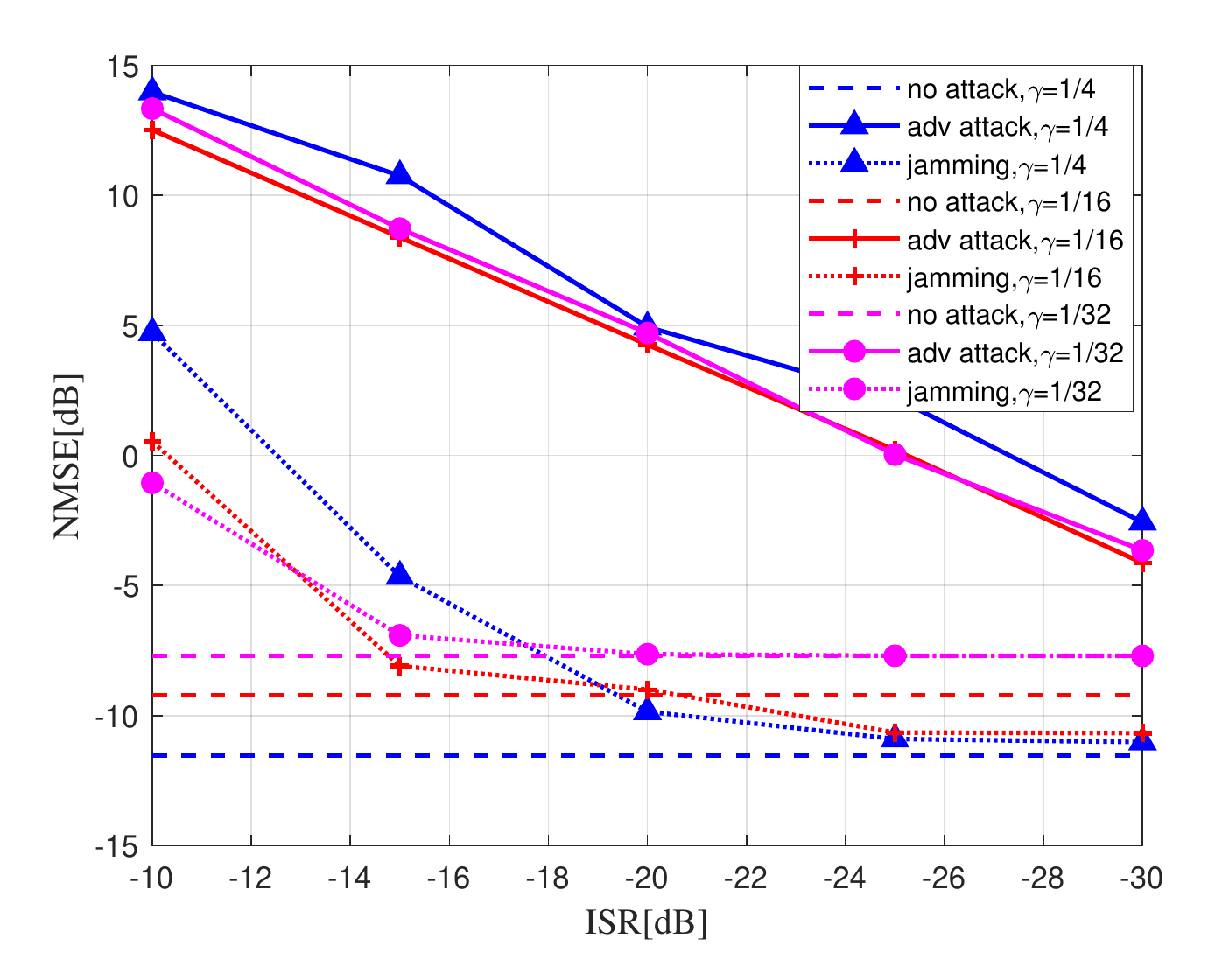}
	\caption{NMSE of CsiNet trained in different compression rates versus SNR with adversarial attack for the indoor scenario.}
	\label{fig:4}
\end{figure}

Previous experiments are performed at using only 1/4 as compression rate, but CsiNet is designed with more than one compression rate to deal with different scenarios. We extend our experiments by attacking models with different compression rates set as \begin{math}1/4\end{math}, \begin{math}1/16\end{math}, and \begin{math}1/32\end{math} using indoor CSI dataset. The results are given in Fig. \ref{fig:4}. The network with lower compression rate owns the NMSE performance under adversarial attack, which is closer to the baseline. Simultaneously, the NMSE performance of the network with lower compression rate drops quickly to the baseline while under jamming attack. Results show that the network with lower compression rate exhibit slight superiority in resisting adversarial and jamming attacks. We interpret this phenomenon as the nature of reconstruction network. A reconstruction work relies on representations that are extracted from original input and trained parameters. CsiNet is forced to rely more on model parameters rather than inputs while less representations are extracted. Hence, CsiNet is less sensible to the variation of the latent representations.

\begin{figure}[t]
	\centering
	\includegraphics[scale=0.63]{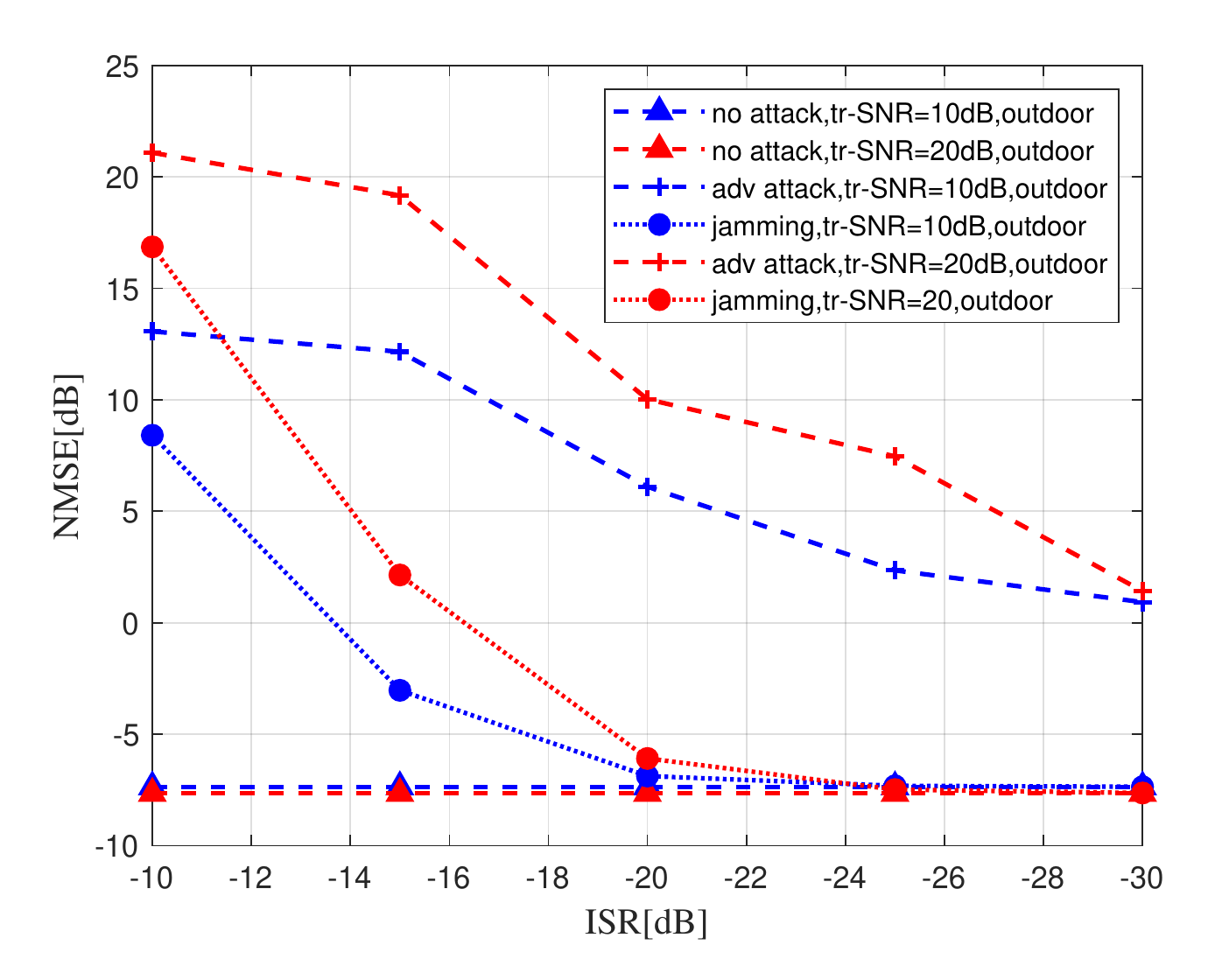}
	\caption{NMSE of CsiNet trained for outdoor scenario versus SNR with adversarial attack with $\gamma$ set to be 1/4.}
	\label{fig:5}
\end{figure}

Considering the complexity of practical channel state, we conduct experiments in indoor and outdoor CSI scenarios to study the effect of CSI complexity on attack performance. We train CsiNet with indoor and outdoor CSI datasets in the AWGN channel, where SNR values set to 10 and 20 dB. We consider the model trained in \begin{math}1/4\end{math} compression rate as example and compare the NMSE performance of CsiNet trained for different scenarios under adversarial and jamming attacks. Fig. \ref{fig:5} shows that NMSE of CsiNet trained with outdoor CSI dataset appears severely influenced by adversarial attack compared to jamming attack. In summary, CSI feedback suffers from great threats under various circumstances, which arouse our attention on real state environments where DL-based physical layer would be exposed under severe threats from a malicious attacker.

\section{Conclusion}
We found that the safety of DL-based CSI feedback against random noise can be guarded by considering noise during training. However, adversarial perturbation still endangered CsiNet despite certain precaution. Due to the broadcast nature of wireless communication, transmitted data can be easily tampered with adversarial perturbation by malicious attackers. With our work, we hope to raise concerns about the security of DL-based physical layer. Further studies in ultra-secure communication system are highly needed.

\begin{thebibliography}{10}
	\providecommand{\url}[1]{#1}
	\csname url@samestyle\endcsname
	\providecommand{\newblock}{\relax}
	\providecommand{\bibinfo}[2]{#2}
	\providecommand{\BIBentrySTDinterwordspacing}{\spaceskip=0pt\relax}
	\providecommand{\BIBentryALTinterwordstretchfactor}{4}
	\providecommand{\BIBentryALTinterwordspacing}{\spaceskip=\fontdimen2\font plus
		\BIBentryALTinterwordstretchfactor\fontdimen3\font minus
		\fontdimen4\font\relax}
	\providecommand{\BIBforeignlanguage}[2]{{%
			\expandafter\ifx\csname l@#1\endcsname\relax
			\typeout{** WARNING: IEEEtran.bst: No hyphenation pattern has been}%
			\typeout{** loaded for the language `#1'. Using the pattern for}%
			\typeout{** the default language instead.}%
			\else
			\language=\csname l@#1\endcsname
			\fi
			#2}}
	\providecommand{\BIBdecl}{\relax}
	\BIBdecl
	
	\bibitem{1}
	P.~Yang, Y.~Xiao, M.~Xiao, and S.~Li, ``6G wireless communications: Vision and
	potential techniques,'' \emph{IEEE Netw.}, vol.~33, no.~4, pp. 70-75, Jul./Aug.
	2019.
	
	\bibitem{2}
	T.~{Wang} et al., ``Deep
	learning for wireless physical layer: Opportunities and challenges,''
	\emph{China Commun.}, vol.~14, no.~11, pp. 92-111, Nov. 2017.
	
	\bibitem{3}
	Z.~Qin, H.~Ye, G.~Y. Li, and B.-H.~F. Juang, ``Deep learning in physical layer
	communications,'' \emph{IEEE Wireless Commun.}, vol.~26, no.~2, pp. 93-99,
	2019.
	
	\bibitem{4}
	H.~Ye, G.~Y. Li, and B.-H. Juang, ``Power of deep learning for channel
	estimation and signal detection in OFDM systems,'' \emph{IEEE Wireless
		Commun. Lett.}, vol.~7, no.~1, pp. 114-117, Feb. 2018.
	
	\bibitem{5}
	T.~{O'Shea}, J.~Corgan, and T.~C. Clancy, ``Convolutional radio
	modulation recognition networks,'' in \emph{Proc. Int. Conf.
		Eng. Appl. Neural Netw.}\relax, pp. 213-226, 2016,
	
	\bibitem{6}
	C.-K.~{Wen}, W.-T.~{Shih}, and S.~{Jin}, ``Deep learning for massive MIMO CSI
	feedback,'' \emph{IEEE Wireless Commun. Lett.}, vol.~7, no.~5, pp. 748-751,
	Oct. 2018.
	
	\bibitem{7}
	I.~J. Goodfellow, J.~Shlens, and C.~Szegedy, ``Explaining and harnessing
	adversarial examples,'' \emph{arXiv preprint arXiv:1412.6572}, 2014.
	
	\bibitem{8}
	M.~Sadeghi and E.~G. Larsson, ``Adversarial attacks on deep-learning based
	radio signal classification,'' \emph{IEEE Wireless Commun. Lett.}, vol.~8,
	no.~1, pp. 213--216, Feb. 2018.
	
	\bibitem{9}
	M.~Sadeghi and E.~G. Larsson, ``Physical adversarial attacks against end-to-end autoencoder
	communication systems,'' \emph{IEEE Commun. Lett.}, vol.~23, no.~5, pp.
	847-850, 2019.
	
	\bibitem{10}
	S.~Sabour, Y.~Cao, F.~Faghri, and D.~J. Fleet, ``Adversarial manipulation of
	deep representations,'' \emph{arXiv preprint arXiv:1511.05122}, 2015.
	
	\bibitem{11}
	J.~{Guo}, C.~{Wen}, S.~{Jin}, and G.~Y. {Li}, ``Convolutional neural network
	based multiple-rate compressive sensing for massive MIMO CSI feedback:
	Design, simulation, and analysis,'' \emph{IEEE Trans. Wireless Commun.}, pp.
	1-1, 2020.
	
	\bibitem{12}
	G.~Caire, N.~Jindal, and M.~Kobayashi, ``Achievable rates of MIMO downlink
	beamforming with non-perfect CSI: a comparison between quantized and analog
	feedback,'' in \emph{Proc. IEEE Asilomar Conf. on Signals Systems
		and Comp.}\relax, pp. 354--358, Oct.-Nov. 2006.
	
	\bibitem{13}
	C.~Zhu, R.~H. Byrd, P.~Lu, and J.~Nocedal, ``Algorithm 778: L-BFGS-B: Fortran
	subroutines for large-scale bound-constrained optimization,'' \emph{ACM
		Trans. Math. Software}, vol.~23, no.~4, pp. 550-560,
	1997.
	
	\bibitem{14}
	P.~Tabacof, J.~Tavares, and E.~Valle, ``Adversarial images for variational
	autoencoders,'' \emph{arXiv preprint arXiv:1612.00155}, 2016.
	
	\bibitem{15}
	L.~Liu, C.~Oestges, J.~Poutanen, K.~Haneda, P.~Vainikainen, F.~Quitin,
	F.~Tufvesson, and P.~De~Doncker, ``The COST 2100 MIMO channel model,''
	\emph{IEEE Wireless Commun.}, vol.~19, no.~6, pp. 92-99, Dec. 2012.
	
\end{thebibliography}
\end{document}